# Assessment Formats and Student Learning Performance: What is the Relation?

**Khondkar Islam, Pouyan Ahmadi, Salman Yousaf**
George Mason University, Fairfax, USA
{kislam2, pahmadi, syousaf3}s@gmu.edu

***Abstract***: *Although compelling assessments have been examined in recent years, more studies are required to yield a better understanding of the several methods where assessment techniques significantly affect student learning process. Most of the educational research in this area does not consider demographics data, differing methodologies, and notable sample size. To address these drawbacks, the objective of our study is to analyse student learning outcomes of multiple assessment formats for a web-facilitated in-class section with an asynchronous online class of a core data communications course in the Undergraduate IT program of the Information Sciences and Technology (IST) Department at George Mason University (GMU). In this study, students were evaluated based on course assessments such as home and lab assignments, skill-based assessments, and traditional midterm and final exams across all four sections of the course. All sections have equivalent content, assessments, and teaching methodologies. Student demographics such as exam type and location preferences are considered in our study to determine whether they have any impact on their learning approach. Large amount of data from the learning management system (LMS), Blackboard (BB) Learn, had to be examined to compare the results of several assessment outcomes for all students within their respective section and amongst students of other sections. To investigate the effect of dissimilar assessment formats on student performance, we had to correlate individual question formats with the overall course grade. The results show that collective assessment formats allow students to be effective in demonstrating their knowledge.*

## Introduction

Assessment and student learning performance are two inseparable elements in higher education. Assessment plays an important role in shaping students' way of thinking towards their learning. The progress of engineering education relies on assessment to a great extent. One of the reasons why assessment has had little impact on the teaching-learning process is due to the absence of studies on the link between assessment type and student performance. A good understanding of this association would lead to mature assessment methods that could evidently enhance student-learning process.

Typically, each course has well-defined learning objectives. Due to several constraints such as class size, restricted time, etc., summative assessments are employed in many universities (Malapati & Murthy, 2013). This type of assessment allows instructors to gauge students' proficiencies in several courses. In summative assessments which is a common method in engineering programs, the instructor generally uses different assessment methods like written exams, papers, projects, presentations and quizzes in order to have a comprehensive understanding of students' learning process. Besides, students have the opportunity of improving their skills throughout the course.

The design of acknowledged practices of instructive assessment have a substantial influence on the advancement of engineering programs and the performance evaluation of students. Imprecise assessments lead to unproductive results, whereas good quality assessments allow instructors to achieve the course outcomes. Also, as it is stated in the research study of Meyers and Nutly (2008), assessment is the first place that students look at in a distinct curriculum. This requires profound analysis of different assessment methods.

The employment of the alleged Bologna process in European universities has revolutionized the structure of teaching, learning approaches and also assessment methods (Flores & Veiga Simão 2015). This paradigm shift suggests an academic rearrangement focusing on both university instructors, students and new assessment methods. Previous empirical works have revealed the significance of assessment and its crucial impact on students' learning (Black and Wiliam 1998; Knight, 2008; Stassen and Kathryn, 2001; Struyven et al., 2005; Olds, Moskal, & Miller, 2005).

In (Almatrafi, Islam & Johri, 2015), authors consider the rate of accessing the material and the time duration used in assessments. They conclude that there is a relation between students' use of learning management system (LMS), Blackboard (BB) and student performance, but there was no proof of remarkable variance across a variety of demographic characteristics. The aim of Pérez et al. (2017) research study was to establish a set of measures to assess one of the most crucial engineering skill, problem solving capability. Gibbs and Simpson (2005) focused on the way assessment arrangements affect student learning. By referencing to both hypothetical and empirical verifications they propose a framework under which assessment supports learning. This framework can be used by teachers to analyse the efficiency of their own assessment practice.

The performance of students across different assessment methods and courses using correlation analysis, is examined in (Malapati & Murthy, 2013). The results of this study confirm that experimental findings when appropriately understood, can benefit educators in planning and fine-tuning the course to enhance student learning. Also, another assessment methodology, the use of correlation analysis is described in the work of (Gomes, Correia & Abreu, 2016). To locate the best exam question for assessing programming knowledge, they devised a game which is played by students in the first class. According to their findings, it is feasible to evaluate student difficulty levels and programming skills by means of dissimilar forms of questions. Likewise, Sean et al. (2007) examined the effects of conducting an initial test in different formats with regard to the way students received corrective feedback. They concluded that regular quizzes with feedback may be more productive in enhancing student learning than taking a multiple choice (MC) test.

Another study (Shankararaman & Gottipati, 2015) has been conducted to offer an automatic alignment technique in which assessments are associated with competencies to determine the consistency between them. Fletcher et al. (2011) research study is an attempt to answer to a seemingly simple question: whether faculty and students have different conceptions of assessments or not, and if so, what kind of implications this would cause in higher education policies? Authors in (Poza-Lujan, Calafate, Posadas-Yagüe & Cano, 2016) also investigate the relation between instructor workload, different evaluation strategies and student grade. Their findings show that continuous evaluation improves student grades whereas exhaustive continuous evaluation (adding written test at the end of each unit) is likely to impose an unnecessary load to instructors without having a noteworthy influence on student grades.

Although diverse assessment approaches have been scrutinized many times recently, more studies are required to gain a better understanding of the abundant methods in which assessment practice affects student learning process. In addition, most of the works have been done in this area, suffer from scholarly flaws such as insignificant sample size, lack of demographic comparisons, and substantial differences in course materials and assessments.

The aim of this paper is to address these drawbacks by comparing student learning outcomes of in-person and asynchronous online web-assisted sections of a core networking undergraduate course with regard to varied assessment methods. We also factored in demographics data in our analysis for all sections. Furthermore, students' preferred assessment is considered as another important source of information on the essence of connotation between assessment and learning. This study is based on previous research conducted at George Mason University to understand the effectiveness of assessments by discovering students' experiences with several dissimilar evaluation methods in multiple sections of the core course.

## Research Study

This empirical study was carried out in the Department of Information Sciences and Technology at George Mason University for three in-class sections (001, 002, 004) and one distance learning (DL) section of a Computer Networking core course (IT 341) that was offered in Fall 2016 semester. In the online section, most of the activities such as lecture materials, pre-recorded lectures, home and lab assignments, etc. were achieved online via BB. However, it was mandatory for students in this class to meet to take their midterm and final exams in a proctored classroom on campus. The other three face-to-face (F2F) sections were all presented in class, but used BB to share course resources with students in similar manner as the DL section. The same instructor taught two sections, and different instructors taught the other sections.

Sessions of each section were administered as follows: half of each session was committed to the theoretical concepts of the course, and the second half was dedicated to practicing skill-based lab assignments conducted with Cisco Packet Tracer network simulator. Throughout this course, other than the classic home/lab assignments and written midterm/final exam, there was one skill-based assessment (SA), which comprehensively gauged students' hands-on skills and one online exam which was taken via Cisco Networking Academy (NetAcad) website.

In order to evaluate students' learning performance, we drastically changed the exam structure in each section. Table 1 summarizes the final exam format for different sections. It's worth noting that, all sections had the same home/lab assignments and similar conceptual materials based on a common syllabus. We considered the same exam structure of 25 multiple-choice questions (MCQs), fill-in-the-blank questions and a numerical problem for sections 001 and 004 because the same instructor taught them. Section 002 had 25 MCQs followed by two essay questions, and the DL section had 45 MCQs followed by a simple numerical problem. It is important to state that for each section, we devised dissimilar essay questions and numerical problems. Students in all sections were given equal amount of time to complete their assessments, and were evaluated by their respective instructor using the same grading rubric across all sections. The course grade was used to evaluate students' learning performance.

**Table 1: Final Exam Formats for Different Sections of IT 341**

| Final Exam Format | Section 001 | Section 002 | Section DL | Section 004 |
|---|---|---|---|---|
| MCQs |  | ✓ | ✓ | ✓ |
| Essay questions | ✓ | ✓ | ✓ |  |
| Fill in the blanks |  |  | ✓ |  |
| Numerical problem |  |  | ✓ |  |

A questionnaire was also designed to gather student demographics data including their gender, work status, age, proximity to campus, prior networking experience, and marital status whose results were reported in (Ahmadi et al., 2017). Moreover, students were queried about the type of question (MCQ, short answer, problem, etc.) they prefer for their exam, and whether they like in-class or take-home exams. The research questions which this paper pursues to highlight are:

- How do diverse assessment arrangements affect student learning performance in different sections of one course?
- Is there a relationship between students' preference for a particular type of question and their final grades?
- Can we find one specific assessment that influence students' final grades the most?
- Do students show concrete connotations with one single type of question? Or are they leaning towards amalgamate of different types of questions?
- How do students observe assessment regarding efficiency in terms of different components of their final grades?

## Participants

In total, 115 undergraduate students participated in the study from 4 different sections. Table 2 summarizes the final grade statistics for all the sections.

**Table 2: IT 341 Sections Statistics**

| Statistics | Section 001 | Section 002 | Section DL | Section 004 |
|---|---|---|---|---|
| Count | 32 | 27 | 28 | 28 |
| Minimum Value | 39.85 | 51.43 | 15.37 | 13.37 |
| Maximum Value | 92.33 | 95.20 | 96.87 | 91.43 |
| Range | 52.48 | 43.77 | 81.50 | 78.06 |
| Average | 72.86 | 77.47 | 79.61 | 76.07 |
| Median | 72.34 | 77.78 | 82.41 | 79.23 |
| Standard Deviation | 11.79 | 9.31 | 18.06 | 18.23 |
| Variance | 139.01 | 86.66 | 326.00 | 332.20 |

## Correlation Analysis

The term correlation usually means the relationship between two or more variables, objects, data points etc. In statistics, the term correlation is measured by the correlation coefficient. A correlation coefficient typically shows the linear dependence between two variables. In mathematical notation, correlation coefficient of two data sets/variables equals to their covariance divided by the product of their individual standard deviations. As it is denoted in the below equation, $r$ is the correlation coefficient $S.D(x)$ is the standard deviation of *variable x*, and $S.D(y)$ is the standard deviation of variable *y* (Dalgaard, 2008).

$$r = \frac{Covariance\ (x, y)}{S.D(x)S.D(y)}$$

If the correlation coefficient of two variables is equal to one, then it means that variables are linearly related and if plotted all the data points will lie on the straight line with positive slope. In our analysis, we used Pearson's product-moment correlation method (Dalgaard, 2008) to test if there is a statistically significant relation between two variables of interest. Our null hypothesis is that the correlation between two variables is equal to zero. To further understand the hypothesis testing, the output of one such test is presented in Figure 1. According to this example, the correlation coefficient equals 0.9200831 with p-value less than 0.05. Therefore, we reject the null hypothesis and confirm that, there is a statistically significant relationship between the variables essays score and total final exam score.

```
> cor.test(b$Essays, b$Total.Score)

        Pearson's product-moment correlation

data:  b$Essays and b$Total.Score
t = 16.774, df = 51, p-value < 2.2e-16
alternative hypothesis: true correlation is not equal to 0
95 percent confidence interval:
 0.8648783 0.9532985
sample estimates:
      cor
0.9200831
```

**Figure 1: Pearson's Product-moment Correlation**

## Final Exam Format Study

Based on our analysis in the above section, we can see from Figure 2, final exam has the highest correlation coefficient amongst the other assessments with the value of 0.86. Hence, we can nominate this exam as the best predictor towards the total course score, which can be regarded as a good indicator for the course performance.

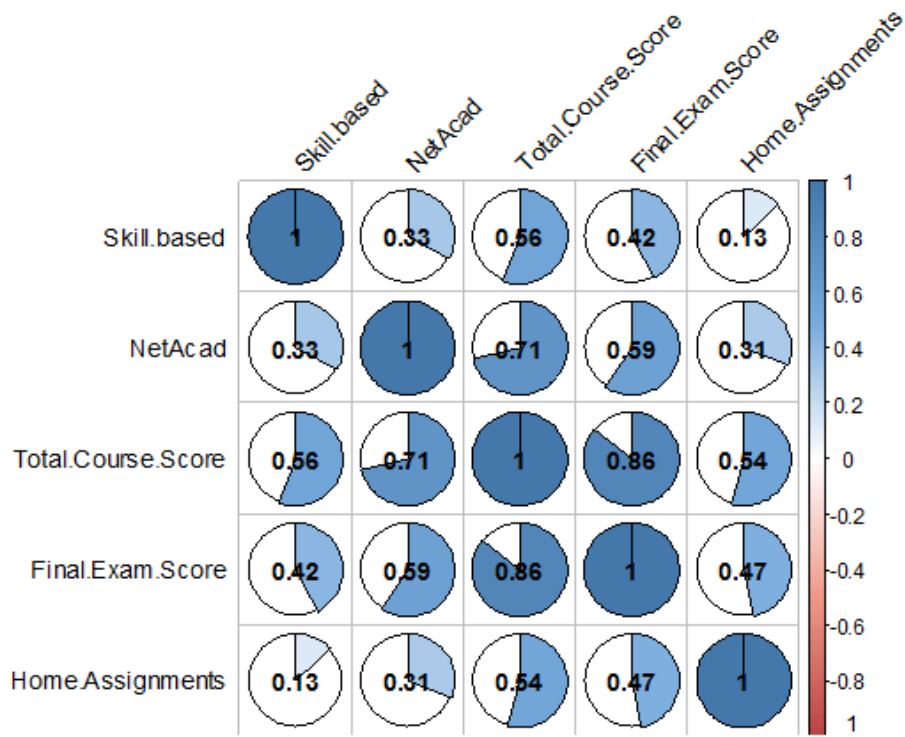

**Figure 2: The Best Total Course Score Predictor Chart**

Since final exam is the best predictor, we examined the impact of individual final exam question format to determine which one is the best indicator for the overall final exam score. To validate our hypothesis, correlation coefficients of final exam question formats are presented in Table 3 where we see that the highest value is for the essay question format. Thus, we can conclude that essay format is the best predictor for the final exam score.

**Table 3: Correlation Coefficients of Final Exam Components**

| Correlation Variables | Correlation Coefficient | p-value | Hypothesis |
|---|---|---|---|
| (Essays, Final Score) | 0.920 | 2.2e-16 | Ha (true correlation is not equal to zero) |
| (MCQs, Final Score) | 0.869 | 4.143e-09 | Ha (true correlation is not equal to zero) |
| (FIB, Final Score) | 0.830 | 1.612e-07 | Ha (true correlation is not equal to zero) |
| (Numerical Problem, Final Score) | 0.430 | 0.0282 | Ha (true correlation is not equal to zero) |

The correlation variables column of Table 3 compares the question format with the final exam score to calculate the correlation coefficient and the corresponding p-values. Figure 3 provides a snapshot of final course grade distribution for all sections.

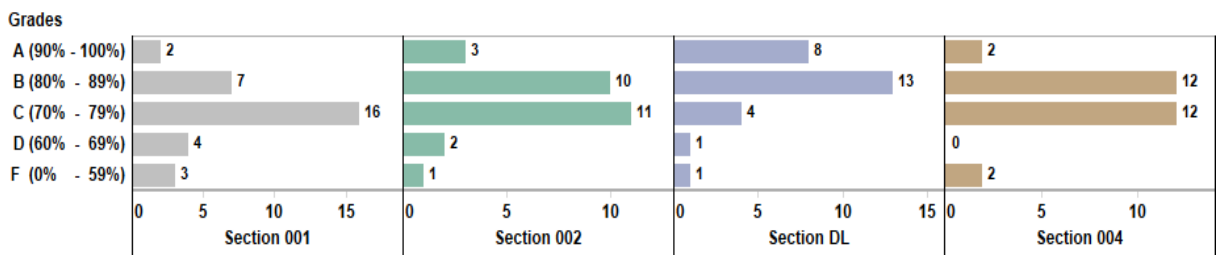

**Figure 3: IT 341 Final Grade Distribution**

According to Figure 3, section 001 was the only section that had the essay question format in the final exam. Based on the results of Figure 2 and respective discussion above, we now turn to the final grade distribution of Figure 3 in which we find student performance of section 001 to be the worst. This validates our findings that the final exam is the best indicator for the overall course grade because section 001 used the essay questions for its final exam.

Figure 4 illustrates the median value of each section that is represented by the horizontal line in each box. The median is based on the final exam score of respective sections. To recap from Table 1 the assessment formats of the final exam in each section: section 001 used essay format only; section 002 had MCQ and essay formats; section DL included MCQ, essay, FIB and numerical problem formats; and section 004 used MCQ format. Section DL final exam median is the highest and section 001 is the lowest. From our correlation analysis done in this paper, the poor performance of section 001 lives up to our expectation that this section would be the worst. Section DL did the best since it used a combination of all four-assessment formats where the students had the opportunity to demonstrate their knowledge via problem solving and objective questions, in addition to the subjective essay questions.

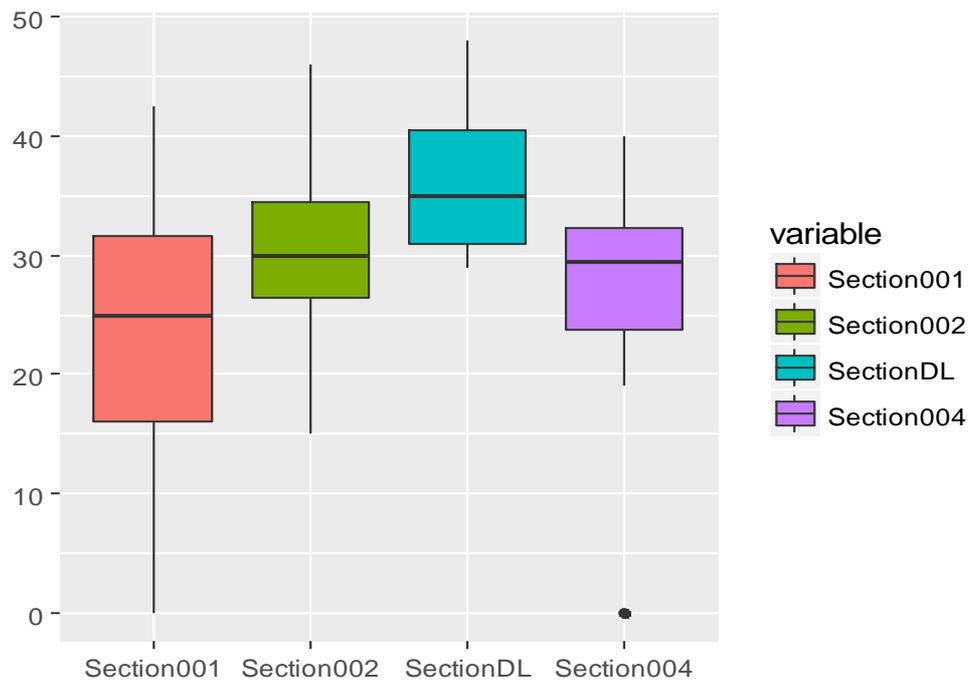

**Figure 4: IT 341 Sections Mean Value Comparison for Final Exam Score**

Moreover, the demographics analysis of Ahmadi et al. (2017) show students tend to lean toward objective questions over subjective essay formats. This is perhaps because with MCQ format, students are able to see the answer options with the question that allows them a better chance of choosing the right answer via process of elimination. This assertion is reinforced from our findings in this study.

## Conclusion and Future Works

In this paper, we presented a novel method to find the most effective assessment format to evaluate student learning performance. A significant amount of data was collected from the learning management system (LMS), Blackboard (BB) Learn that made it possible for us to perform extensive statistical manipulation via correlation analysis. Our findings demonstrate that final exam assessment format is the best predictor to determine the course outcome for a core networking course in our department. Since the performance of section 001 was the worst and it used only essay question format in the final exam, we were able to conclude that this format is the best predictor for the final exam score. Section DL did the best since it used a combination of all four-assessment formats where the students had the opportunity to demonstrate their knowledge via problem solving and objective questions, in addition to the subjective essay questions. The demographics survey that was administered in all sections presented in our previous work also support our findings since students expressed their preference towards combined assessment techniques over a single format.

For our future work, we will investigate the relationship between student navigational behaviour and their learning performance while they access course content in Blackboard LMS.